\documentclass[twocolumn]{el-author}

\usepackage{epstopdf}
\usepackage[normalem]{ulem}

\newcommand{\ua}{\uparrow}
\newcommand{\nc}{\newcommand}
\makeatletter
\newcommand{\vast}{\bBigg@{1.75}}
\newcommand{\Vast}{\bBigg@{2.4}}
\newcommand{\vastl}{\bBigg@{4}}
\newcommand{\Vastl}{\bBigg@{5}}
\makeatother
\nc{\da}{\downarrow} \nc{\hc}{\hat{c}} \nc{\hS}{\hat{S}}
\nc{\bra}{\langle} \nc{\ket}{\rangle} \nc{\eq}{equation (\ref}
\nc{\h}{\hat} \nc{\hT}{\h{T}}\nc{\be}{\begin{eqnarray}}
\nc{\ee}{\end{eqnarray}}\nc{\rd}{\textrm{d}}\nc{\e}{eqnarray}\nc{\hR}{\hat{R}}\nc{\Tr}{\mathrm{Tr}}
\nc{\tS}{\tilde{S}}\nc{\tr}{\mathrm{tr}}\nc{\8}{\infty}\nc{\lgs}{\bra\ua,\phi|}\nc{\rgs}{|\ua,\phi\ket}
\nc{\hU}{\hat{U}}\nc{\lfs}{\bra\phi|}\nc{\rfs}{|\phi\ket}\nc{\hZ}{\hat{Z}}\nc{\hd}{\hat{d}}\nc{\mD}{\mathcal{D}}
\nc{\bd}{\bar{d}}\nc{\bc}{\bar{c}}\nc{\mc}{\mathcal}\nc{\ea}{eqnarray}\nc{\mG}{\mathcal{G}}\nc{\bce}{\begin{center}}
\nc{\ece}{\end{center}}
\date{28th April 2018}

\begin{document}

\title{Effective throughput of MISO systems over $\kappa-\mu$ shadowed fading channels: MGF based analysis}

\author{H. Al-Hmood and H. S. Al-Raweshidy}

\abstract{The effective throughput of multiple-input single-output (MISO) wireless communication systems over $\kappa-\mu$ shadowed fading channels is analysed. To obtain exact closed-from expression, the moment generating function (MGF) of the instantaneous signal to noise ratio (SNR) with independent and identically distributed (i.i.d.) transmit antennas is employed. In addition, a special case of integer values for the fading parameter is performed to simplify the derived expression. The asymptotic behaviour of the effective throughput at high values of SNR is also studied. The Monte Carlo simulations and the numerical results are presented to verify the validation of our analysis.}

\maketitle

\section{Introduction}
The effective throughput (capacity, rate) has been proposed to measure the performance of the wireless communication systems the quality of service (QoS) constraints, such as system delays, that have not been taken into consideration by Shannon [1]. Accordingly, the study of the effective throughput over generalised fading channels has been given a special attention by the researchers. This is because these channels include most of the class models such as Nakagami-$m$ but they provide close results to the practical measurements. For example, in [2], the analysis is investigated over $\kappa-\mu$ shadowed fading channel which is a composite of $\kappa-\mu$ and Nakagami-$m$ distributions that represent the multipath with line-of-sight (LoS) scenario and shadowing impacts, respectively. The parameters $\kappa$ and $\mu$ stand for the powers ratio between the domination and the scattered waves components and number of multipath clusters, respectively.
\par More recent, the effective throughput of multiple-input single-output (MISO) systems over fading channels have been widely presented in the open technical literature. For instance, the analysis over independent and identically (i.i.d.) $\kappa-\mu$ fading channels is given in [3]. The $\eta-\mu$ and $\alpha-\mu$ fading conditions that are proposed to model the non-LoS (NLoS) and non-linearity communication scenarios are used in [4], and [5], respectively. A unified expression of effective throughput using the moment generating function (MGF) of the instantaneous signal to noise ratio (SNR) of MISO wireless communications system is derived in [6] with applications to the generalised $K$ which is composite of Nakagami-$m$/log-normal and Weibull/gamma fading scenarios. The analysis over $\kappa-\mu$ shadowed fading channels with i.i.d. transmit antennas is presented in [7]. However, the provided performance metric, namely, the effective throughput, is included double infinite series. Thus, an approximation has been suggested by the authors via employed the Gamma distribution to obtain simple closed-from expressions.     
\par In this letter, unlike [7], the MGF approach is used to express the effective throughput over i.i.d. $\kappa-\mu$ shadowed fading channels in simple exact closed-from. To this effect, the derived result is given in terms of the extended generalised bivariate Meijer's $G$-function (EGBMGF) that is not available in the most popular mathematical software packages such as MATLAB and MATHEMATICA. However, several works are included codes to computes this function. For further simplification to the expression of the effective throughput, a special case when both $\mu$ and $m$ are integer numbers with $m \geqslant \mu$ is considered. In addition, the asymptotic analysis at high SNR is also studied to obtain a clear insight for the behaviour of the effective throughput under this condition. 
\section{System model}
For MISO systems, the normalised effective capacity over $\textbf{h}$ channel is expressed as [3, eq. (3)]
\label{eqn_1}
\begin{equation}
\mathcal{R}(\rho, \theta)=-\frac{1}{A}\mathrm{log}_2\bigg(\mathbb{E}\vast{\{}\vast(1+\frac{\rho}{L}\mathbf{h h^\dagger}\vast)^{-A}\vast{\}}\bigg)
\end{equation}
where $A \triangleq \theta TB/\mathrm{ln}2$ with $\theta$, $T$, and $B$ stand for the delay exponent, block duration, and bandwidth, respectively, and $\mathbb{E}\{.\}$ denotes the expectation. 
\section{The MGF of $\kappa-\mu$ shadowed fading}
The MGF of the instantaneous SNR, $\gamma$, over $\kappa-\mu$ shadowed fading is given as [8, eq.(5)]
\label{eqn_2}
\begin{equation}
\mathcal{M}_\gamma(s)={\mathcal{A}^\mu \mathcal{B}^m} \frac{(s+\mathcal{A})^{m-\mu}}{(s+\mathcal{A B})^m} 
\end{equation}
where $\mathcal{A} = \frac{\mu(1+\kappa)}{\bar{\gamma}}$, $\mathcal{B} = \frac{m}{\mu \kappa+m}$, $\bar{\gamma}$ is the average SNR, and  $m$ is the shadowing severity parameter.
\par For $L$ i.i.d. transmit antennas, the MGF is expressed as
\label{eqn_3}
\begin{equation}
\mathcal{M}_\gamma(s)= {\mathcal{M}_\gamma^L}(s)
\end{equation}
\par Accordingly, the MGF over $L$ i.i.d. $\kappa-\mu$ shadowed fading channels is yielded as
\label{eqn_4}
\begin{equation}
\mathcal{M}_\gamma(s)= ({\mathcal{A}^\mu \mathcal{B}^m})^L \frac{(s+\mathcal{A})^{L(m-\mu)}}{(s+\mathcal{A B})^{Lm}} 
\end{equation}
\section{Effective capacity of MISO $\kappa-\mu$ shadowed fading}
According to [6], (1) can be rewritten in terms of the MGF as follows
\label{eqn_5}
\begin{equation}
\mathcal{R}(\rho, \theta)=-\frac{1}{A}\mathrm{log}_2 \bigg(\frac{1}{\Gamma(A)}\int_0^\infty s^{A-1} e^{-s}\mathcal{M}_\gamma\vast(\frac{\rho}{L}s\vast)d{s}\bigg)
\end{equation}
\par Now, inserting (4) in (5) and performing some mathematical manipulations, this yields
\label{eqn_6}
\begin{align}
\mathcal{R}(\rho, \theta)&=-\frac{1}{A}\mathrm{log}_2 \bigg(\frac{1}{\Gamma(A)} \nonumber\\
&\times \int_0^\infty s^{A-1} e^{-s} \vast(1+\frac{\rho}{\mathcal{A} L}s \vast)^{L(m-\mu)} {\vast(1+\frac{\rho}{\mathcal{A B} L}s \vast)^{-Lm}} d{s}\bigg)
\end{align}
\par With the help of the identities [9, eq. (11)] and [9, eq. (10)], the functions exp($-a$) and $(1+a)^b$ can be, respectively, expressed in terms of the Meijer G-function as follows
\label{eqn_7}
\begin{align}
e^{-s}&=G^{1,0}_{0,1} \Vast[\begin{matrix}
  -\\\\
  0\\
\end{matrix}\Vast\vert s
\Vast]\nonumber\\
\vast(1+\frac{\rho}{\mathcal{A} L}s \vast)^{L(m-\mu)}&=\frac{1}{\Gamma(L(\mu-m))}
G^{1,1}_{1,1} \Vast[\begin{matrix}
    1+L(m-\mu)\\\\
  0\\ 
\end{matrix}\Vast\vert \frac{\rho}{\mathcal{A} L}s
\Vast]\nonumber\\
\vast(1+\frac{\rho}{\mathcal{A} \mathcal{B} L}s \vast)^{-Lm}&=\frac{1}{\Gamma(L m)}
G^{1,1}_{1,1} \Vast[\begin{matrix}
    1-Lm\\\\
  0\\ 
\end{matrix}\Vast\vert \frac{\rho}{\mathcal{A} \mathcal{B} L}s
\Vast]
\end{align}
where $G^{.,.}_{.,.}[.]$ is the Meijer's G-function [9, eq. (5)].
\par Substituting (7) in (6), we obtain 
\label{eqn_8}
\begin{align}
\mathcal{R}(\rho, \theta)&=-\frac{1}{A}\mathrm{log}_2 \Vast(\frac{1}{\Gamma(A) \Gamma(L(\mu-m)) \Gamma(L m)}\nonumber\\ 
&\times \int_0^\infty s^{A-1} G^{1,0}_{0,1} \Vast[\begin{matrix}
  -\\\\
  0\\
\end{matrix}\Vast\vert s
\Vast] 
G^{1,1}_{1,1} \Vast[\begin{matrix}
    1+L(m-\mu)\\\\
  0\\ 
\end{matrix}\Vast\vert \frac{\rho}{\mathcal{A} L}s
\Vast] \nonumber\\ 
& G^{1,1}_{1,1} \Vast[\begin{matrix}
    1-Lm\\\\
  0\\ 
\end{matrix}\Vast\vert \frac{\rho}{\mathcal{A} \mathcal{B} L}s
\Vast]
ds\Vast)
\end{align}
\par Using [10, eq. (9)] to compute the integral in (8) in exact closed-form. Thus, this yields
\label{eqn_9}
\begin{align}
\mathcal{R}(\rho, \theta)&=-\frac{1}{A}\mathrm{log}_2 \Vast(\frac{1}{\Gamma(A) \Gamma(L(\mu-m)) \Gamma(L m)}\nonumber\\ 
&\times G^{0,1:1,1:1,1}_{1,0:1,1:1,1} \Vast[\begin{matrix}
    1-A\\\\
  -\\ 
\end{matrix}\Vast\vert
\begin{matrix}
    1+L(m-\mu)\\\\
  0\\ 
\end{matrix}\Vast\vert
\begin{matrix}
    1-Lm\\\\
  0\\ 
\end{matrix}\Vast\vert
\frac{\rho}{\mathcal{A} L}, \frac{\rho}{\mathcal{A} \mathcal{B} L}
\Vast]\Vast)
\end{align}
\par One can see that (9) is expressed in terms of the EGBMGF that is not yet implemented in MATLAB and MATHEMATICA software packages. Therefore, a MATHEMATICA code that is given in [10] is utilised in this letter. 
\section{Effective capacity of MISO $\kappa-\mu$ shadowed fading with integer $\mu$ and $m$}
When both $\mu$ and $m$ are integer numbers, i.e., $\mu$ and $m$ $\in \mathbb{Z}^+$ and under the condition $m\geqslant \mu$, the following identity can be employed [11, eq. (1.111)]
\label{eqn_10}
\begin{equation}
 \vast(1+\frac{\rho}{\mathcal{A} L}s \vast)^{L(m-\mu)}=\sum_{j=0}^{L(m-\mu)} {{L(m-\mu)}\choose{j}}\vast(\frac{\rho}{\mathcal{A}L} s \vast)^j
\end{equation}
where ${{b}\choose{a}} = \frac{b!}{a! (b-a)!}$ is the binomial coefficient. 
\par Plugging (10) in (6), we have
\label{eqn_11}
\begin{align}
\mathcal{R}(\rho, \theta)&=-\frac{1}{A}\mathrm{log}_2 \bigg(\frac{1}{\Gamma(A)} \sum_{j=0}^{L(m-\mu)} {{L(m-\mu)}\choose{j}}\vast(\frac{\rho}{\mathcal{A}L} \vast)^j\nonumber\\
&\times \int_0^\infty e^{-s} s^{A+j-1} {\vast(1+\frac{\rho}{\mathcal{A B} L}s \vast)^{-Lm}} d{s}\bigg)
\end{align}
\par After using the solution of the above integral in (11) that is available in [3] and doing some mathematical simplifications, this yields
\label{eqn_12}
\begin{align}
\mathcal{R}(\rho, \theta)&=-\frac{1}{A}\mathrm{log}_2 \bigg(\sum_{j=0}^{L(m-\mu)} {{L(m-\mu)}\choose{j}}\bigg(\frac{L}{\rho}\bigg)^{A}  \bigg(\frac{\mathcal{B}^A}{\mathcal{A}}\bigg)^{j}\nonumber\\
&\times (A)_j U\vast(A+j;A+j-Lm;\frac{\mathcal{A B} L}{\rho}\vast)\bigg)
\end{align}
where $(.)_j$ is the Pochhammer symbol and $U(.;.;.)$ is the Tricomi hypergeometric function defined in [12, eq. (07.33.02.0001.01)].
\section{Asymptotic analysis at high SNR} 
Although the expressions in (9) and (12) are derived in exact closed-form, they do not provide a clear insight when the parameters of the system and the channel are changed. Therefore, the asymptotic analysis at high SNR is deduced from (6) after plugging $\rho \rightarrow \infty$ and performing some straightforward mathematical operations. Accordingly, this yields 
\label{eqn_13}
\begin{align}
\mathcal{R}(\rho \rightarrow \infty, \theta)\thicksim -\frac{1}{A}\mathrm{log}_2 \bigg(\frac{ \Gamma (A-L \mu)}{\Gamma(A)}
 \vast(\frac{\mathcal{A} L}{\rho} \vast)^{L\mu} \mathcal{B}^{Lm}\bigg)
\end{align}

\begin{figure}[h]
\centering
  \includegraphics[width=3.5 in, height=2.4 in]{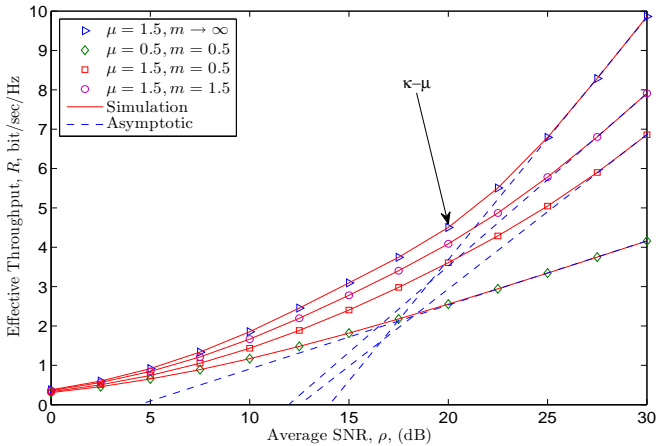} 
\centering
\caption{Simulated, numerical, and asymptotic effective throughput versus the average SNR, $\rho$, of MISO over i.i.d. $\kappa-\mu$ shadowed fading channels for different values of $\mu$, $m$, $\kappa = 3$, $\bar{\gamma} = -5$ dB, $A = 2$, and $L = 2$.}
\end{figure} 
\begin{figure}[h]
\centering
\includegraphics[width=3.5 in, height=2.4 in]{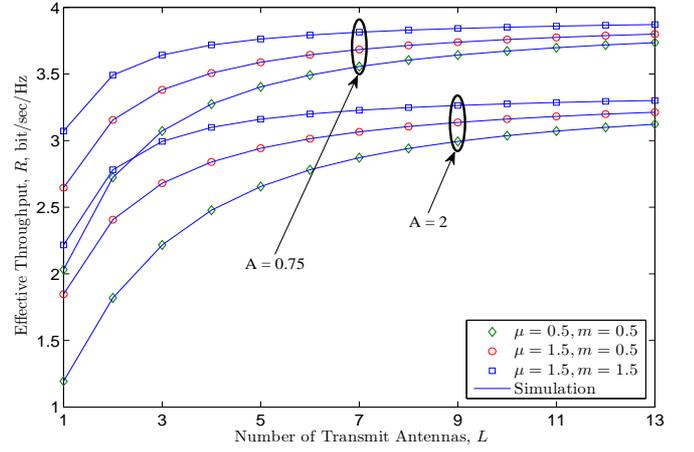} 
\centering
\caption{Simulated and numerical effective throughput versus the number of the transmit antennas, $L$, of MISO over i.i.d. $\kappa-\mu$ shadowed fading channels for different values of $\mu$, $m$, $\kappa = 3$, $\bar{\gamma} = -5$ dB, $A = 2$, and $\rho = 15$ dB.}
\end{figure}
\section{Numerical results} 
Fig. 1 and Fig. 2 demonstrate the Monte Carlo simulations and the numerical results of the effective throughput of MISO systems over i.i.d. $\kappa-\mu$ shadowed fading channels versus the SNR, $\rho$, and number of transmit antennas, $L$, respectively, for different values of $\mu$ and $m$. In both figure, the parameters of the Monte Carlo simulations that are derived by averaging the results over 10$8$ realizations are $\bar{\gamma} = -5$ dB, and $\kappa = 3$. As expected, the effective throughput increases with the increasing in the value of $\mu$ or/and $m$. This refers to the large number of multipath clusters that arrives at the receiver in the former and to the low effect of the shadowing fading in the latter. Moreover, the asymptotic behaviour of the effective throughput at high SNR is also explained in Fig. 1. In addition, the performance becomes better when the number of the transmit antennas increase as a result of improving in the total values of $\kappa$, $\mu$, and $m$.     

\section{Conclusion}
The MGF approach was employed to analyse the effective throughput of MISO over i.i.d. $\kappa-\mu$ shadowed fading scenarios. To this end, exact closed-form expression in terms of the extended generalised bivariate Meijer's $G$-function (EGBMGF) was derived. This expression was further simplified via assuming $\mu$ and $m$ are integer numbers. The asymptotic analysis of the effective throughput at high SNR was also given in order to obtain a good insight about the behaviour of the system that can not be investigated by the exact model. Its noteworthy that the derived results can be utilised to study the effective capacity of MISO over various fading channel models such as Nakagami-$m$ and $\kappa-\mu$ after inserting the fading parameters, namely, $\kappa$, $\mu$, and $m$ for a certain value.

\vskip3pt
\copyright
\vskip5pt

\noindent Hussien Al-Hmood (\textit{Electrical and Electronics Engineering Department, University of Thi-Qar, Thi-Qar, Iraq})\\
\noindent E-mail: hussien.al-hmood@{brunel.ac.uk, eng.utq.edu.iq}\\
H. S. Al-Raweshidy (\textit{Electronic and Computer Engineering Department, College of Engineering, Design and Physical Sciences, Brunel University London, UK})

\end{document}